\documentclass[journal]{IEEEtran}
\usepackage{ifpdf}
\usepackage{cite}
\usepackage{graphicx}
\usepackage{booktabs}
\usepackage[cmex10]{amsmath}
\usepackage{amsfonts}
\usepackage{algorithmic}
\usepackage{array}
\usepackage{mdwmath}
\usepackage{mdwtab}
\usepackage{eqparbox}
\usepackage{multirow}
\usepackage{cleveref}
\usepackage{mathtools}
\usepackage{url}
\usepackage{ragged2e}
\usepackage[table,svgnames]{xcolor}
\usepackage{xcolor}
\usepackage{balance}

\begin{document}
%
\title{Analysing and Measuring the Performance of Memristive Integrating Amplifiers}
%
%
%

\author{Jiaqi Wang,
        Alexander Serb, Christos Papavassiliou, Sachin Maheshwari,
        Themistoklis Prodromakis 
\thanks{J. Wang, A. Serb, S. Maheshwari and T. Prodromakis are with the Centre for Electronics Frontiers, Zepler institute, University of Southampton, UK, SO171BJ. C. Papavassiliou is with the Department of Electrical and Electronic Engineering, Imperial College London, UK. Corresponding author e-mail: (jw9y17@soton.ac.uk).}
}

\maketitle

\begin{abstract}
Recording reliably extracellular neural activities is an essential prerequisite for the development of bioelectronics and neuroprosthetic applications. Recently, a fully differential, 2-stage, integrating pre-amplifier was proposed for amplifying and then digitising neural signals. The amplifier featured a finely tuneable offset that was used as a variable threshold detector. Given that the amplifier is integrating, the DC operating point keeps changing during integration, rendering traditional analysis (AC/DC) unsuitable. In this work, we analyse the operation of this circuit and propose alternative definitions for validating the necessary key performance metrics, including: gain, bandwidth, offset tuning range and offset sensitivity with respect to the memory states of the employed memristors. The amplification process is analysed largely through investigating the transient behaviour during the integration phase. This benchmarking approach is finally leveraged for providing useful insights and design trade-offs of the memristor-based integrating amplifier.
\end{abstract}


%
\IEEEpeerreviewmaketitle

\section{Introduction}

\IEEEPARstart{R}{ecording} neural signals using implantable microsystems is essential to the development of diagnostic and therapeutic solutions \cite{d_rodrigues_2017}, Brain Machine Interfaces (BMIs) \cite{lebedev2006brain} and neuroscience research \cite{lee_lee_kiani_jow_ghovanloo_2010}. The implantable device typically contains electrodes as well as front-end and back-end module, where raw neural signals collected from electrodes will be fed into the other two modules for further processing \cite{patil2016implantable}. After processing, analogue neuronal trains\cite{1052457} or digital format \cite{hashemi2020multi} will be transmitted to external devices wirelessly. With digital output, a neural spike (Action Potential, AP) detection algorithm which comprises threshold detection and digitisation can be applied in back-end stage typically \cite{harrison_2003}. For an implantable device, this is required to have low power/heat dissipation ($<80\, mW / {cm}^2$) in order to avoid damaging surrounding tissue \cite{PMID:9881955}. The low power dissipation contributes to high integration density. Furthermore, both dc offset \cite{4120899} and minute extracellular neural activity signals (in the order of 10s-100s of $\mu V$) picked up by electrodes will be fed into front-end devices for amplification and filtering \cite{vijay_obien_franke_frey_hierlemann_2019}. In summary, the implantable front-end module needs to have low-power dissipation, low-noise and also to reject dc offset and other noise interference.

To achieve low power consumption, a number of multi-channel neural recording architectures has been proposed \cite{5256307}\cite{4956983}\cite{7489028}. It is clear that the energy consumed in the analogue multiplexer before ADC can be reduced to improve power efficiency. From the  system level point of view,  Serb et al.\cite{serb_prodromakis_2017} propose to perform spike detection and digitisation directly on the neural signal from electrodes in order to save power from processing local field potentials (LFP) which will be discarded.

Preamplifiers are critical for boosting the extremely weak input signals to levels where they can be further processed and so they act as the first stage in any neural recording processing (a result of the Friis formula) \cite{4511242}. Alternatively, the operational transconductance amplifier-capacitor (OTA-C) structure is suitable for bio-electronic devices as the low-pass filter for neural signals \cite{1347327}. The objective of combining an OTA with load capacitor is to integrate signals instead of simply amplifying them in continuous mode in order to boost effective gain. A different technique has been proposed to compensate the DC offset of the electrode-tissue interface \cite{4039585}\cite{6418430}\cite{6679691}. The Harrison topology is capable of rejecting large dc offset, operates in continuous mode and is the current standard in the field \cite{Harrison2003}. It is possible to conduct threshold detection directly on the signals in Harrison amplifier. \cite{serb_prodromakis_2017}. 

With the characteristic of analogue modulation of their resistive state, memristive devices can be utilised in CMOS circuit as trimming component \cite{stathopoulos2017multibit}. Such an integrating pre-amplifier enhanced with offset tuning for ultra-fine threshold detection was proposed \cite{serb_prodromakis_2017}. In this work, memristive devices were utilised as non-volatile resistive loads \cite{waser_aono_2007} to trim the offset voltage with high precision. \par

The architecture and preliminary analysis presented in \cite{serb_prodromakis_2017} demonstrated the general operating principle of what we may describe as a "memristive integrating amplifier". In this work we add detail on the operation of this type of amplifier as well as investigate how important parameters such as clocking and differential/common input voltage affect performance. One of the challenges identified in doing this is defining important performance (e.g. gain, bandwidth, CMRR and etc.) in a manner suitable to the operation of integrating amplifiers. We provide such metrics that suit the particular implementation of the memristive integrating pre-amplifier. The mathematical descriptions of the resulting metrics and insight obtained from examining the behaviour of transistors during the key integration phase of the amplifier illuminate various trade-offs that characterise the design. This work has been done using commercially available $0.18 \mu m$ CMOS technology with $1.8V$ supply voltage across all the experiments. \par

The paper is organised as follows: A brief overview of the pre-amplifier and its operation, followed by the re-definition of its key performance metrics is presented in section \ref{secBack}. Simulation set-up, analysis and results are shown in section \ref{secRes}. A discussion of design trade-offs and other points of interest pertaining to the amplifier design is in section \ref{secDis}. Finally, section \ref{secConc} summarises and concludes the paper.


\section{Fundamental operation and analysis}\label{secBack}


\subsection{Amplifier design overview}
\begin{figure}[!t]
\centering
\includegraphics [width=9cm]{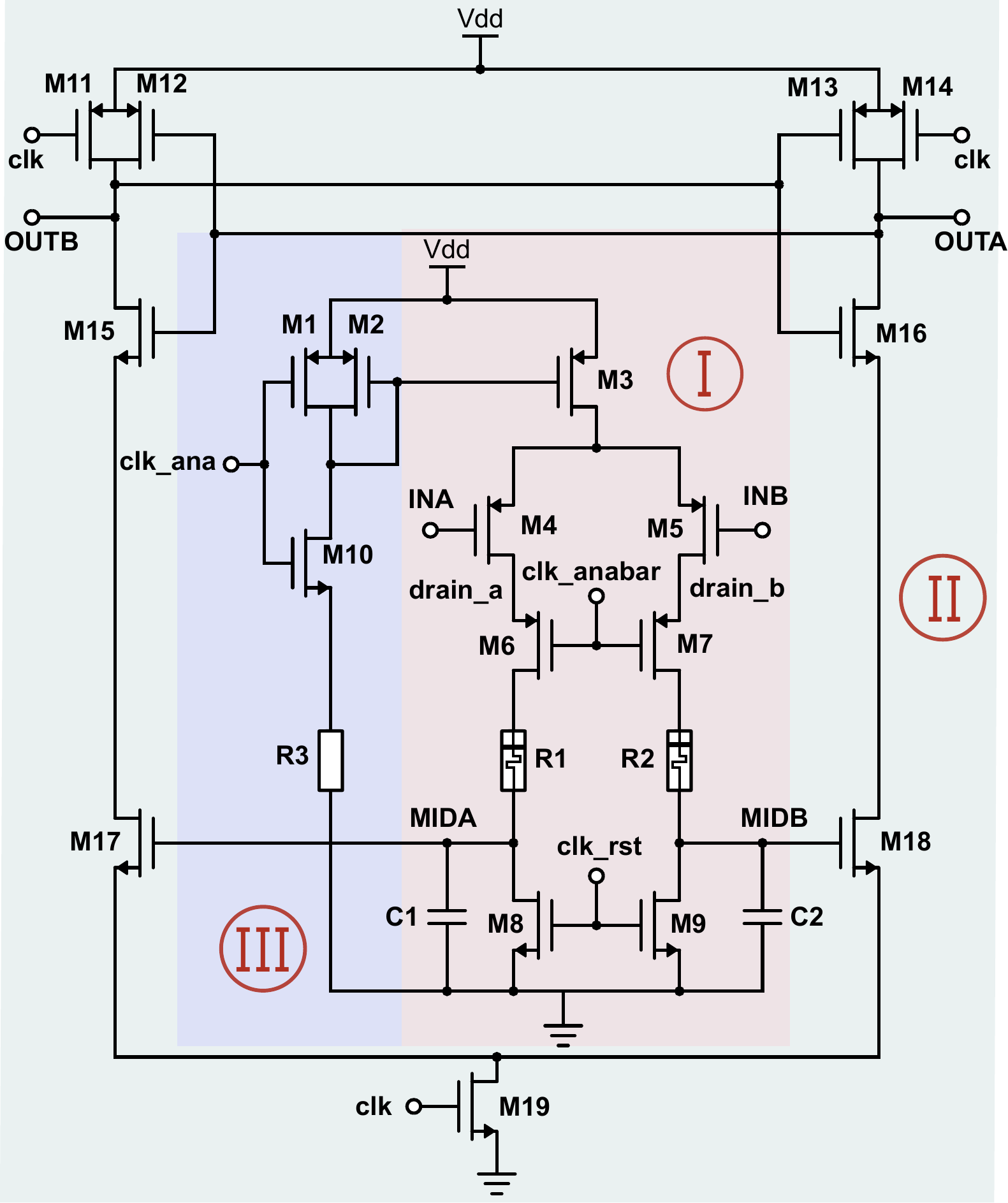}
\caption{Architecture of the simplified pre-amplifier. (a) The circuit can be divided into three parts: (I) core integrating amplifier, (II) dynamic latch comparator, (III) current bias control unit. In this paper, the clocking signals are  all assumed to be generated by a shared source and be strictly periodic.}
\label{fig_1}
\end{figure}

\begin{figure}[!t]
\centering
\includegraphics [width=8cm]{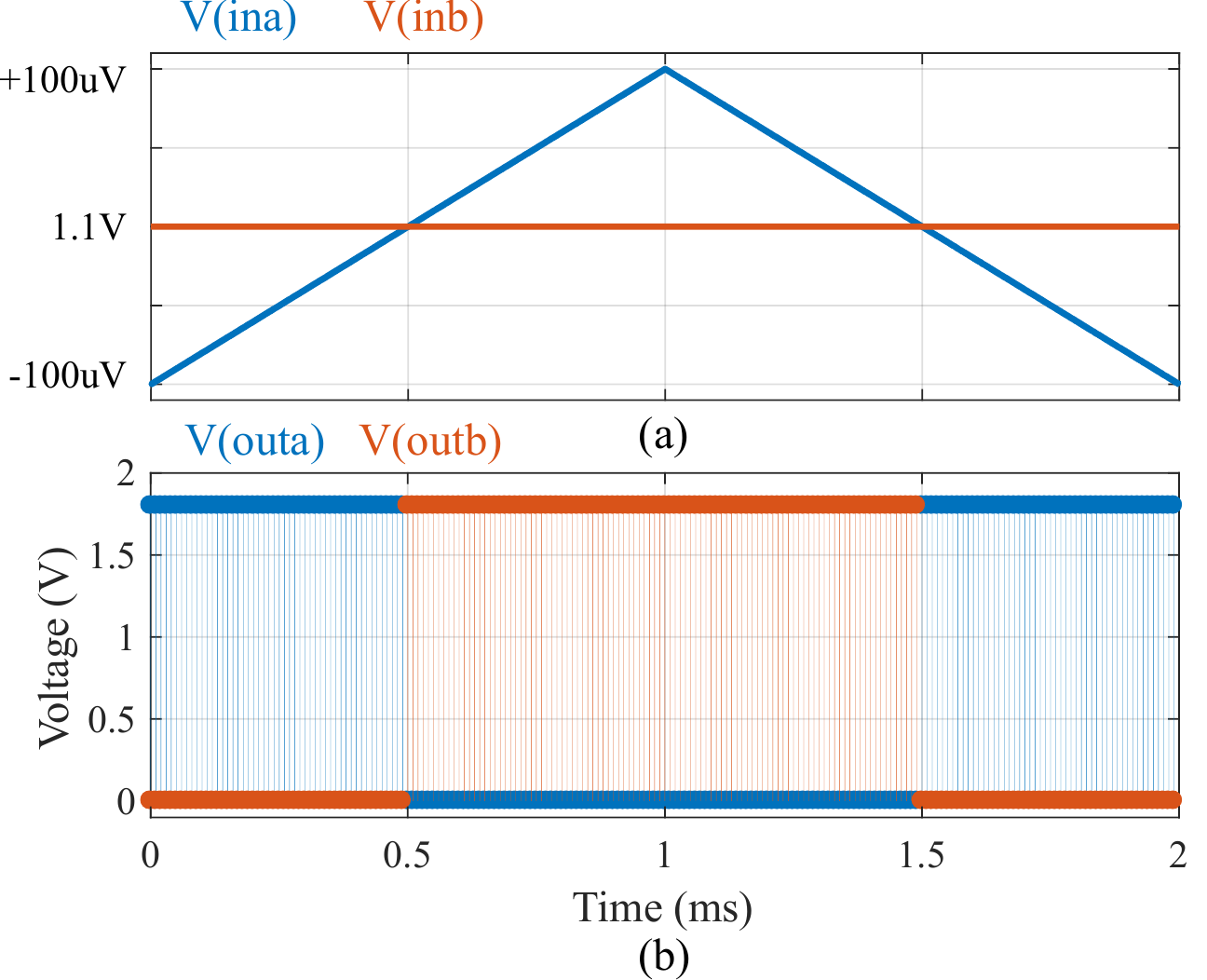}
\caption{Pre-amplifier basic functionality test: Input A (ina) is slowly swept between [$1.1V - 100\mu V$, $1.1V + 100\mu V$] over $2ms$ while the pre-amplifier is carrying out a conversion every $10\mu s$ to detect the relationship between inputs A and B. Input B (inb) remains stable at $1.1\,V$ throughout. In this test the amplifier was balanced ($R1 = R2$). When $V_{ina}<V_{inb}$, the left branch current is larger than the right branch current, inducing $V_{mida} - V_{midb}>0$. The DLC captures this relation and generates binary signals: $V_{outa} = 1$ and $V_{outb} = 0$, which appears in the bottom trace as a predominantly orange output trace. Conversely when $V_{ina}>V_{inb}$, $V_{outa} = 0$ and $V_{outb} = 1$, which appears as a combined orange/blue output trace. Note: this type of simulation can also be used to test the offset tuning range and tuning sensitivity on the resistive state of memristive devices. When $R1>R2$, $V_{ina}$ must be lower than $V_{inb}$ to ensure a balanced output, creating an offset. This is read in the output trace as an encroachment of the blue region into the orange (and vice versa for $R1<R2$).}
\label{fig_2}
\end{figure}

The architecture studied is a modified/simplified version of the original design proposed in \cite{serb_prodromakis_2017}; it is shown in Fig. \ref{fig_1}. It consists of three main sections: I) a fully differential core amplifier (effectively a single-stage analogue amplifier acting as the 1st stage of the design), II) a dynamic latched comparator (DLC) amplifying and quantising the output state of the core amp  and III) a current bias unit powering the system's core. The overall system operates as a threshold detection circuit which compares two minute input signals and ultimately outputs a binary flag, as shown in Fig.\ref{fig_2}.

\begin{figure}[!t]
\centering
\includegraphics [width=8cm]{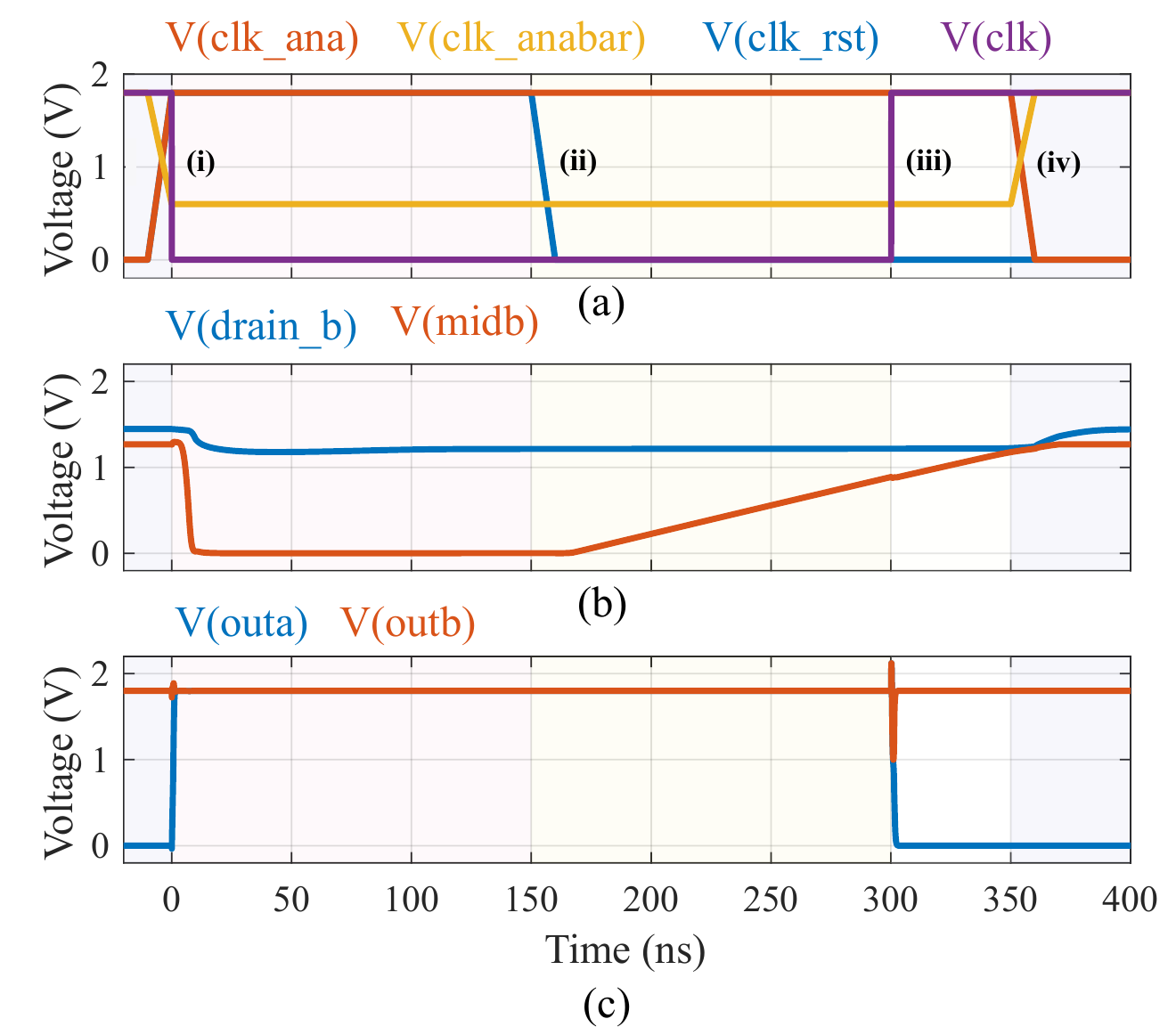}  
\caption{Timing diagram for neural signal detection. The timing diagram is captured from one detection cycle, where we set this cycle starts from $0ns$. The signal-detection cycle period was set at $350\,ns$, which is subdivided into four operational phases: (i)reset, (ii) integrating, (iii) digitisation and (iv) off phases. Top panel: clocking scheme (see schematic in Fig. \ref{fig_1}). Middle panel: drain signal of input transistor M5 ($drain\_b$) and integrating node voltage ($midb$). Bottom panel: digital output signals; it is these signals that generate the bottom panel of Fig. \ref{fig_2}. (In this simulation, $ina = 1.1V + 50\mu V$ and $inb = 1.1V$)}
\label{fig_3}
\end{figure}

Each threshold detection operation is carried out in four phases that we label as: (i) reset, (ii) integrating, (iii) digitisation and (iv) off. These  are illustrated in Fig. \ref{fig_3}. They remain unchanged from the original work and act as follows:

In the reset phase (i) the core amplifier is on ($clk\_ana$, $clk\_rst$: high, $clk$, $clk\_anabar$: low) and the load capacitors are discharged ($V_{mida/b} = 0$), so that voltage/current in core amplifier is initialised and cleared before integration commences in the next phase.

In the integrating phase (ii) ($clk\_ana$: high, $clk\_anabar$, $clk\_rst$, $clk$: low) the reset transistors (M8\&M9) are switched off and the currents flowing through the branches of the core amplifier drain into the load capacitors. From a `large signal' perspective, $V_{mida}$ and $V_{midb}$ continuously increase during integration. In terms of `small signal', $\Delta V_{mida-midb}$ increases with time and normal operation is maintained so long as the cascode transistors M6\&M7 remain in the saturation region. The voltage difference between nodes $mida$ and $midb$ is impacted by the charging speed/current and integration time. Memristors R1\&R2 work as trimming devices and tune the offset of the core with very high sensitivity ($1\mu V/k\Omega$ shown in the original paper). At the end of this phase, $V_{mida/b}$ should be high enough to successfully trigger the DLC and $\Delta V_{mida-midb}$ should be as large as possible for maximising gain.
    
In the digitisation phase (iii) ($clk\_ana$, $clk$: high, $clk\_anabar$, $clk\_rst$: low) $clk$ goes high, triggering the DLC to perform the conversion of $V_{mida/b}$ into the final digital outputs. By convention we take the output from the branch where output `1' represents a spike while `0' represents the absence of a spike. Shortly after the decision is committed by the DLC, the core amplifier is turned off as the system re-enters the off phase.

Finally, in the off phase (iv) ($clk$: high, $clk\_ana$, $clk\_anabar$, $clk\_rst$: low), the tail current is cut off by setting $clk\_ana$ to zero. The pre-amplfier is turned off and stops recording neural signals. $clk\_anabar$ is also deactivated (goes to high) thus preventing the accumulated charge across the large gate capacitances of M4\&M6 from draining away.

\subsection{Differences vs. the original design}
The design under study has been simplified vis-a-vis the original from \cite{serb_prodromakis_2017} as follows: First  a fixed clocking scheme was adopted. The previous design featured an asynchronous clock generation circuit embedded in each channel. When it determined that the result would be available on nodes $mida/midb$ it triggered, on-demand, the clocks. However, it has been observed that the integration results for the very small differential input signals of interest always become available at fixed intervals; therefore, an on-demand triggering system is not required here. Second, the sizes of the input transistor pair  were decreased.
In \cite{serb_prodromakis_2017}, huge transistors were introduced as input pair to reduce noise. However, the associated large parasitic capacitances together with the $\mu A$ tail currents result in a low transit frequency ($f_T$). Therefore, in consideration of speed and power consumption the sizes of input transistors have been decreased.

\begin{table}[h!]
    \centering
    \caption{Sizes of devices in the proposed architecture, where the bias current of core amplifier is $I_{tail} = 3\, \mu A$. $R3$ is replaced by a diode connected NMOS. The supply voltage is $1.8V$ and control signals are designed to full range swing except $clk\_ anabar$ swings between $0.6V$ and $1.8V$.}
    \begin{tabular}{c c c c}
    \hline
    Devices & W/L ($\mu m$) & Devices & W/L ($\mu m$) \\
    \hline
    M1, M2, M3 & 3/3 & M11-M14 & 2/0.6 \\
    M4, M5 & 200/1 &  M15,M16 & 1/0.6\\
    M6, M7 & 20/1 & M17, M18 & 2/0.6\\
    M8, M9 & 1/0.6 &  M19 & 1/0.6 \\
    M10, R3 & 5/1 & C1, C2 & 200 $fF$\\
    \hline
    \end{tabular}
    \label{tab:sizing}
\end{table}

\subsection{Key performance metrics}
The main performance indicators for the core amplifier include: gain, bandwidth, offset tuning range and sensitivity on memristor resistance, noise performance, input range, common-mode rejection ratio (CMRR) and power consumption. All these metrics (with the exception of power) mostly depend on the integrating phase, when amplification is conducted. In this stage the cascode transistors are in saturation mode. As $V_{mida/b}$ keep increasing throughout the integration phase, there is no set DC operating point. nonetheless, because this is an extremely small signal amplifier, the current flowing through each branch is under normal circumstances approximately the same and constant. This allows for analysis similar in spirit to regular small-signal analysis by using transient simulations for obtaining the relevant data. Standard DC operating point and AC analysis cannot be applied here directly. It is perhaps more appropriate to think of $V_{mida/b}$ as `large signal' in the $mV$-range and $\Delta V_{mid} = V_{mida}-V_{midb}$ as `small signal' in the $\mu V$-range.

\subsubsection{Gain}
The gain is defined, as usual, as the ratio of the output signal amplitude over the input signal amplitude,  $\delta V_{out}/\delta V_{in}$. For the core amplifier this translates into $\delta V_{mid}/\delta V_{in}$, where $\delta V_{mid}$ is taken at the end of the integration phase and $\delta V_{in}$ is considered constant for the purposes of this analysis.

A more explicit formula can be obtained for the gain: The input differential voltage induces  through the input differential pair and its associated current branches a difference in currents:

\begin{equation}\label{st1}
    \delta i = \delta V_{in} \cdot g_m
\end{equation}
where $g_m$ is the transconductance of the input differential pair in normal operating conditions. This induces a difference of charge on the load capacitors:

\begin{equation}\label{st2}
    \delta Q = \delta i \cdot \tau
\end{equation}
where $\tau$ is the integration phase duration. Finally, this gets transformed into the voltage difference we observe at $\delta V_{mid}$ through the load capacitances $C$:

\begin{equation}\label{st3}
    \delta V_{mid} = \delta Q/C
\end{equation}

Combining the above yields the gain (G):

\begin{equation}\label{st4}
    G = \frac{\delta V_{mid}}{\delta V_{in}} = \frac{g_m \cdot \tau}{C}
\end{equation}

Given that we know that the currents filling each load capacitor are approximately constant and equal we can express integration time $\tau$ as a function of the difference in $V_{mid}$ levels at the start and end of integration $\Delta V_{mid} = V_{mid}|_{t=t_e} - V_{mid}|_{t=t_0}$, where $t_0$ and $t_e$ denote the start and end of the integration phase. In our case $V_{mid}|_{t=t_0} = \text{GND} = 0$ and therefore $\Delta V_{mid} = V_{mid}|_{t=t_e}$. This is a voltage level that we can adjust by choosing appropriate values for the tail current of the amplifier and the integration time. Given this interdependence we now seek to find an expression for $\tau$ that depends only on engineering parameters. We begin by observing that:

\begin{equation}\label{St1}
    \Delta V_{mid} = \Delta Q/C
\end{equation}
where $\Delta Q$ is the total charge accumulated on each node ($mida/b$) as a result of the tail current. This can, however, be easily expressed as:

\begin{equation}\label{St2}
    \Delta Q \approx i_{tail/2} \cdot \tau
\end{equation}
where $i_{tail/2}$ is the half-tail current of the amplifier core. This allows us to express $\tau$ as follows:

\begin{equation}\label{St3}
    \tau = \Delta V_{mid} \cdot C / i_{tail/2}
\end{equation}
where we replaced the approximation symbol with an equality for clarity, since the deviation is expected to be sufficiently small under normal operation.

Now we can substitute Eq.\ref{St3} into Eq. \ref{st4} and obtain gain as:

\begin{equation}\label{res1}
    G = \frac{g_m \cdot \Delta V_{mid}}{i_{tail/2}}
\end{equation}
which further simplifies to:

\begin{equation}\label{res2}
    G = TE \cdot \Delta V_{mid}
\end{equation}
where $TE$ is the transconductor efficiency factor of the input diff pair transistors. In other words the differential gain of the pre-amplifier core only depends on the $TE$ and the voltage range over which we are integrating. Integration time and tail current can be freely traded off, in principle (but consider noise, etc.). Note that $\Delta V_{mid}$ represents voltage difference during the integration phase, while $\delta V_{mid}$ is the output that captured at the end of integration.

\subsubsection{Bandwidth}
In an integrating amplifier, such as the one studied here, the notion of bandwidth is somewhat different than in continuous mode systems because the output is not a continuous waveform whose Fourier component at some frequency can be compared in magnitude to an input stimulus of the same frequency. Instead, our amplifier output is a single value that is influenced (in magnitude) by the input in proportion to the input's absolute integral. For a unit magnitude pure tone signal of angular frequency $\omega = 2\pi f$, the maximum absolute integral within a time window $2a$ is given by: 

\begin{equation}\label{Ieffmain}
    I_{eff,max}(\omega) = \left|\int_{-a}^{a} cos(\omega t)dt\right| = \left|\frac{2sin(\omega a)}{\omega}\right| \leq \left|\frac{2}{\omega}\right|
\end{equation}
where $I_{eff}$ stands for 'effective integral'. There is no need for introducing a phase shift $\phi$ into $cos(\omega t)$; using the trignomoetric identity for cosine of sum of angles we can easily prove that $I_{eff}$ maximises for $\phi = 0$.

At DC the integral is simply $2a$ and subsequently it decreases within the envelope of $1/f$ as frequency increases. If we divide $I_{eff,max}$ by the length of the window we obtain what can be interpreted as an attenuation factor:

\begin{equation}\label{lambdaf}
    \lambda(\omega) = \left|\frac{sin(a\omega)}{a\omega}\right| = \left|sinc(a\omega)\right| \leq \left|\frac{1}{a\omega}\right|
\end{equation}

\begin{figure}[!t]
\centering
\includegraphics [width=8.5cm]{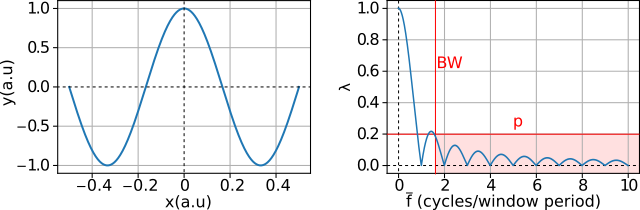}
\caption{Illustration of bandwidth definition within the context of the integrating amplifier. Left panel: Illustration of a pure tone wave fitting exactly 1.5 times within a time window of length 1. Right panel: Attenuation factor (Eq. \ref{lambdaf}) as a function of tone frequency $\overline{f}$ in units of cycles/window period ($\frac{2\pi}{\tau} = \frac{\pi}{a}$). An indicative bandwidth value (BW) is shown for $p=20\%$. $\lambda$ declines with $\frac{1}{a\omega}$.}
\label{Iefffig}
\end{figure}

Fig. \ref{Iefffig} illustrates the evolution of $\lambda$ with $\overline{f}$; the frequency of the sinusoid in units of $\frac{\pi}{a}$. We many now define the effective bandwidth of the amplifier as the frequency above which $\lambda(\overline{f})$ is always below a certain value $p$. An indicative measure of bandwidth may be given by $p = 20\%$. For an integration period of $1\, \mu s$ this yields around $1.5 \, MHz$ bandwidth. Naturally, $p$ can be set to another suitably chosen value to yield different appropriate bandwidth figures. This metric holds only so long as the resulting frequency is much lower than all other RCs in the amplifier core, and thus we have no additional attenuation. It is worth noting that the most typical neural signals of interest, action potentials (spikes), last in the order of $ms$. This implies that the even spike features of the order of 100s of $\mu s$ will be integrated without any significant attenuation.

Throughout our analysis we make the following approximation: the amplifier is integrating linearly throughout its integration voltage range $\Delta V_{mid}$. The input differential signals of interest are so small that linear approximations can be assumed to hold throughout the whole system (\cref{st1,St2}). In practice, there will be some additional distortion due to the changing $V_{ds}$ experienced by the cascode transistors, but we currently ignore this effect in our analysis.

\subsubsection{Tuning Sensitivity and range}
The memristive devices applied in the current branches regulate the charging speed to load capacitors by modulating the effective output resistance of the core amplifier as seen by the capacitive load. To see the mechanics of this action we refer to the schematic in Fig. \ref{fig_1} and the standard equation for the impedance of a drain-degenerated MOSFET, looking into the source. When this is applied to the source of M6 we obtain:

\begin{equation}\label{ZM6}
    Z_{s6} \approx \frac{1}{g_{m6}}\left(1 + \frac{R_1}{R_{o6}}\right)
\end{equation}
where $Z_{s6}$ is the impedance looking into the source of M6, $g_{m6}$ is the differential transconductance of M6, and $R_{o6}$ the output resistance of M6.

Extending this principle to calculate the impedance of M4, as drain-degenerated by the M6-R1 cascade we obtain:

\begin{equation}\label{ZM4}
    Z_{s4} \approx \frac{1}{g_{m4}}\left(1 + \frac{Z_{s6}}{R_{o4}}\right)
\end{equation}
which eventually unfolds to:

\begin{equation}\label{ZM4b}
    Z_{s4} \approx \frac{1}{g_{m4}} + \frac{1}{g_{m6}g_{m4}R_{o4}} + \frac{R_1}{g_{m6}R_{o6}g_{m4}R_{o4}}
\end{equation}

A similar expression also applies for the right current branch.

Setting $A = \frac{1}{g_{m4}} + \frac{1}{g_{m6}g_{m4}R_{o4}}$ and $B = \frac{1}{g_{m6}R_{06}g_{m4}R_{o4}}$ we can express the impedances seen by M3 looking into each current branch as:

\begin{equation}\label{Zl}
    Z_l \approx A + BR_1
\end{equation}
\begin{equation}\label{Zr}
    Z_r \approx A + BR_2
\end{equation}
where $Z_l = Z_{s4}$ is the left current branch impedance and $Z_r$ is the right branch impedance.

Next, examining the distribution of tail current across the branches we obtain an expression for the left branch current $i_l$ as follows:

\begin{equation}\label{iLchain}
    i_l \approx i_T\frac{A+BR_2}{2A + B(R_1+R_2)}
\end{equation}
where $i_T=i_3$ is the tail current. Given that $B \ll 1$ (as it is the product of two maximum FET amplifier gains), $i_l$ can be further approximated as follows:

\begin{equation}\label{iLapprox}
    i_l \approx \frac{i_T}{2}\left(1 - \frac{B}{2A}(R_1 - R_2)\right)
\end{equation}

Similarly for the right branch current $i_r$:

\begin{equation}\label{iR}
    i_r \approx \frac{i_T}{2}\left(1 + \frac{B}{2A}(R_1 - R_2)\right)
\end{equation}

This yields a total current imbalance of:

\begin{equation}\label{di}
    i_l - i_r \approx \Delta i = - i_T \cdot \frac{B}{2A}(R_1 - R_2)
\end{equation}


which if divided by the common transconductance of the input differential pair transistors yields the required voltage offset to rebalance the branches as a function of the difference in RRAM resistive states:

\begin{equation}\label{Vos}
    V_{os} \approx V_{ina} - V_{inb} = \frac{\Delta i}{g_{m4,5}}
\end{equation}
which when fully unfolded yields:

\begin{equation}\label{Vosfull}
    V_{os} \approx - \frac{(R_1 - R_2)i_T}{2R_{o,cas} g_{m,in} (1 + g_{m,cas}R_{o,in})}
\end{equation}
where we have renamed our variables to explicitly stress the common values of output impedances and differential transconductances of the input differential pair and cascode transistors ($R_{o,cas}$ = output impedance of cascode transistor, $g_{m,in}$ = diff. transconductance of the input diff pair).

This result relies on the standard small-signal assumptions that the various $g_m$s and $R_o$s remain constant, all transistors involved remain saturated (either over or below threshold) and crucially, it makes no other assumptions on the voltage present at the load capacitors. So long as: a) the $g_m$s of all transistors remain mostly unchanged and b) the change in load capacitor voltage does not affect the absolute difference in RRAM resistive states seen by the system, the capacitors charge uniformly under balanced conditions ($V_{in} = V_{os}$). Whilst condition (a) can be reasonably approximated as true in saturation, condition (b) is not generally true because of the non-linearity in the $I-V$ of the RRAM devices \cite{messaris2018data}. Analysis of this phenomenon is outside the scope of the paper as it is RRAM technology-specific, but in general if the absolute resistive state difference changes as the integration process progresses, we obtain offset voltage drift that may potentially affect operation when a fixed, non-zero offset is specifically required (e.g. for threshold detection with the offset acting as threshold).

Overall, Eq. \ref{Vosfull} shows that in small-signal conditions the offset voltage of the core amplifier is proportional to the difference in RRAM resistive states divided by the maximum transistor gains of the input diff pair and cascode transistors. This division explains the extreme fineness of tuning achievable.

The tuning range can in principle be extended under the rule of Eq. \ref{Vosfull} for as long as the underlying assumptions hold. We note two important limiting conditions: 1) If the imbalance in currents becomes large, eventually the assumption of equal $g_m$s on both current branches collapses. Exactly when this occurs depends on the tightness of the specifications. 2) If the voltage dropped across the larger of the pair $R_{1,2}$ becomes comparable to the capacitor voltage range through which the amplifier can integrate while maintaining transistor saturation (normal operation), eventually the amplifier will run out of integration voltage headroom. Thus, introducing a headroom vs. maximum tuning range headroom (so long as condition (2) remains the dominant limit).



\subsubsection{Input-Referred Noise}
The amplifier's core noise is dominated by the input differential pair. The reasons are the same as in continuous mode amplifiers such as the Harrison \cite{Harrison2003}: the input pair provides substantial gain through its $g_m$ and thus mitigates the input-referred contributions from downstream elements (primarily the cascode transistors and the RRAM devices).

The standard MOSFET input referred-noise model containing both thermal and flicker noise is given by the following expression for spectral density \cite{lamb1997low}:
\begin{equation}\label{noise}
    \overline{V_{in}^2}(f) = 4kT\gamma \frac{1}{g_m} + \frac{K}{C_{ox}WLf}
\end{equation}
where $k$ is Boltzmann's constant, $T$ is the absolute temperature, $\gamma = \frac{2}{3}$ for long-channel transistors and higher for shorter channel devices, $K$ a typically empirically determined factor scaling $1/f$ noise, $C_{ox}$ the gate capacitance, $W,L$ the transistor sizes and $f$ denotes (linear) frequency.

In our amplifier the noise from each transistor in the input differential pair from Eq. \ref{noise} propagates to the output via the gain $G$ from Eq. \ref{res1} and is then moderated by the attenuation factor $\lambda$ from Eq. \ref{lambdaf}. Moreover, bearing in mind that the amplifier's output is the difference $V_{mid}$ and that it is operating in a `nearly balanced' regime, the total noise spectral density equation at the outputs becomes:

\begin{equation}\label{totnoise}
    \overline{V_{out,total}^2}(f) = 2 \cdot \overline{V_{in}^2}(f) \cdot G^2 \cdot \lambda^2(f)
\end{equation}
where we substitute all $\omega$s with $f$ for simplicity and have assumed that both branches contribute equally to noise.

We note the following: First, the application of $\lambda(f)$ turns white noise into $1/f$ (more accurately $1/af$) and $1/f$ noise into $1/f^2$, as is typical of single-pole low-pass filters. Second, if we desire short integration periods, noise moderation effect by $\lambda(f)$ may become too weak to make any practical difference because of the $a \ll 1$ factor.

Finally, input-referring Eq. \ref{noise} and the contributions of $\lambda(f)$ (which can be ignored in this case), we obtain the following noise profile:

\begin{equation}\label{InputNoise}
    \overline{V_{in,total}^2} = 2 \cdot \lambda^2(f) \left( \frac{8}{3}kT \frac{1}{g_{m}} + \frac{K}{C_{ox}WLf} \right)
\end{equation}
where the $g_m$, $W$ and $L$ factors are the same (at least approximately) for both transistors in the input diff pair.


\subsubsection{Input Range}
Under normal operation, the input differential pair transistors M4,5 must be in subthreshold saturation. This implies two operating conditions: (1) A minimum drain-source voltage $|V_{ds,min}| = m \cdot V_{T}$, where $V_{T}$ is the thermal voltage and good rule of thumb for ensuring subthreshold saturation is $3 \leq m \leq 4$ (here we will use $m=4$) \cite{5404144}. (2) We need an appropriate gate-source voltage $|V_{gs,4}| (< |V_{th}|)$ that allows the transistor to pass $\approx i_{tail}/2$ in subthreshold saturation. This is treated as approximately constant in this analysis.

Therefore the common mode voltage $V_{CM}$ is bounded: The top boundary is simply:

\begin{equation}\label{vcmswingtop}
    V_{DD} - |V_{ds,sat,3}| - |V_{gs,4}| \geq V_{CM}
\end{equation}
where $V_{ds,sat,x}$ is the drain-source saturation voltage of transistor $x$. Exceeding the boundary causes M3 to triode and simultaneously encroaches on $V_{gs,4}$, progressively shutting the amplifier down.

The bottom boundary hinges on maintaining the input diferential pair in subthreshold saturation ($|V_{ds,4}| \geq |V_{ds,min,4}|$):

\begin{equation}\label{lowbound}
    |V_{ds,4}| \approx  (V_{CM} + |V_{gs,4}|) - (V_{anabar,low} + |V_{gs,6}|) \geq 4 \cdot V_{T}
\end{equation}
where $V_{gs,6}$ is the gate-source voltage allowing the cascode transistor to pass $\approx i_{tail}/2$. This is also treated as approximately constant in this analysis. The 2nd term is recognised as $V_{drain\_a}$ under normal operation and node voltage $V_{drain\_a}$ can be seen in the schematic of Fig. \ref{fig_1}. This unfolds to:

\begin{equation}\label{vcmswingbot}
    V_{CM} \geq V_{anabar,low} + |V_{gs,6}| - |V_{gs,4}| + 4 \cdot V_T
\end{equation}

Here, the cascode transistor M6 enforces a specific and relatively fixed value of $V_{drain\_a}$ under the control of $V_{anabar,low}$ (similarly for M7 and $V_{drain\_b}$). Combining Eqs. \ref{vcmswingtop} and \ref{vcmswingbot} we can find the approximate value of $V_{anabar,low}$ above which the input differential pair runs out of common mode range:

\begin{equation}\label{vcmlimit}
    V_{anabar,low} = V_{DD} - |V_{ds,sat,3}| - 4 \cdot V_T - |V_{gs,6}|
\end{equation}

From here we can see the trade-off between common mode and integration voltage ranges (directly connected to gain). If the input stage of the amplifier is AC-coupled, the required $V_{CM}$ range may become very small.

\subsubsection{CMRR and CMGD}
In continuous mode amplifiers CMRR (common more rejection ratio) is defined as the ratio of the differential gain vs. the common mode gain. In our case this is given by:

\begin{equation}\label{CMRReq}
    \text{CMRR} = \frac{A_{dm}}{A_{cm}} = \frac{dG}{dA_{cm}}
\end{equation}
where $A_{dm}, A_{cm}$ are the differential and common mode gains respectively.

In a perfectly balanced amplifier (nominal design) this will be zero at first order, so it would be perhaps more informative to measure this directly in silico.

There is a slightly different effect which will impact our integrating amp and can be analysed easily: Gain distortion vs. common mode voltage $V_{CM}$:

We define this `common mode gain distortion' as:

\begin{equation}\label{CMGD}
    \text{CMGD} = \frac{dG}{dV_{CM}}
\end{equation}

Taking Eq. \ref{res2} and substituting $g_m = \frac{i_{tail/2}}{V_{gs,4}}$ we obtain:

\begin{equation}\label{Ggm}
    G = \frac{\Delta V_{mid}}{V_{gs,4}}
\end{equation}

We can now unfold the derivative $\frac{dG}{dV_{CM}}$ as follows:

\begin{equation}\label{unfold1}
    \frac{dG}{dV_{CM}} = \frac{d\frac{\Delta V_{mid}}{V_{gs}}}{dV_{gs}}\cdot \frac{dV_{gs}}{dV_g} = -\frac{\Delta V_{mid}}{V_{gs}^2} \cdot \frac{dV_{gs}}{dV_g}
\end{equation}
where $\frac{dV_{gs}}{dV_g} \approx 1$ due to the high impedance of M3.

We note that this value could easily be as low as 1 (e.g. consider the case of $\Delta V_{mid} = 0.5V$ and $V_{gs} = 0.7V$). This means that for every Volt of change in $V_{CM}$ the gain deviates by a unit (e.g. $G = 25$ at $V_{CM} = x\,V$ means $G = 26$ at $V_{CM} = (x-1)\,V$). Nevertheless, for indicative values of $G=25$ and $V_{CM}$ fluctuations in the low 100s of $mV$ we obtain gain deviations/errors in the order of $1\%$.





\section {Performance measurements and Results}\label{secRes}
In this section, the suitably defined performance parameters from the previous section will be assessed for an example design in simulation. We split the results into two groups for convenience: differential mode and common mode effects. Under differential mode-related effects we examine the differential gain, bandwidth, and tuneable range/sensitivity of offset vs. RRAM device resistive state. Under `Common mode-related effects' we include input range (largely determine by the common mode by assumption) and CMRR/CMGD. Finally, power consumption is discussed on its own at the end. For these simulations we used a commercially available $0.18 \mu m$ CMOS technology with $\text{VDD} = 1.8V$.

\subsection{Differential Mode Effects}
\subsubsection{Gain}
For the purposes of amplifier gain analysis, we have run multiple, single data-point amplification transients sweeping a range of input differential voltages centred around zero. These simulations are under nominal conditions for this study: no added noise, mismatch or process variation was included.

There are two main experiments: First we set an integration phase run where $\delta V_{in} \neq 0$ and the $clk$ signal does not interrupt the integration process, but rather lets it run its course until both $V_{mida/b}$ saturate. Thus, the important features of the resulting waveform (e.g. position of peaks) are revealed. A key question we seek to answer here is whether there is an optimum time to stop the amplification in order to reliably obtain maximum gain, and if so when that occurs. The second experiment uses a fixed clock allowing us to explore the gain linearity for fixed integration period: we run multiple simulation runs with $\delta V_{in}$ swept from $-100\mu V$ to $100\mu V$ with integration period $\tau = 150ns$. The key question here is whether the amplifier has a usable linear range centred around the $0V$ differential input and if so, how wide it is.

\begin{figure}[!t]
\centering
\includegraphics [width=8cm]{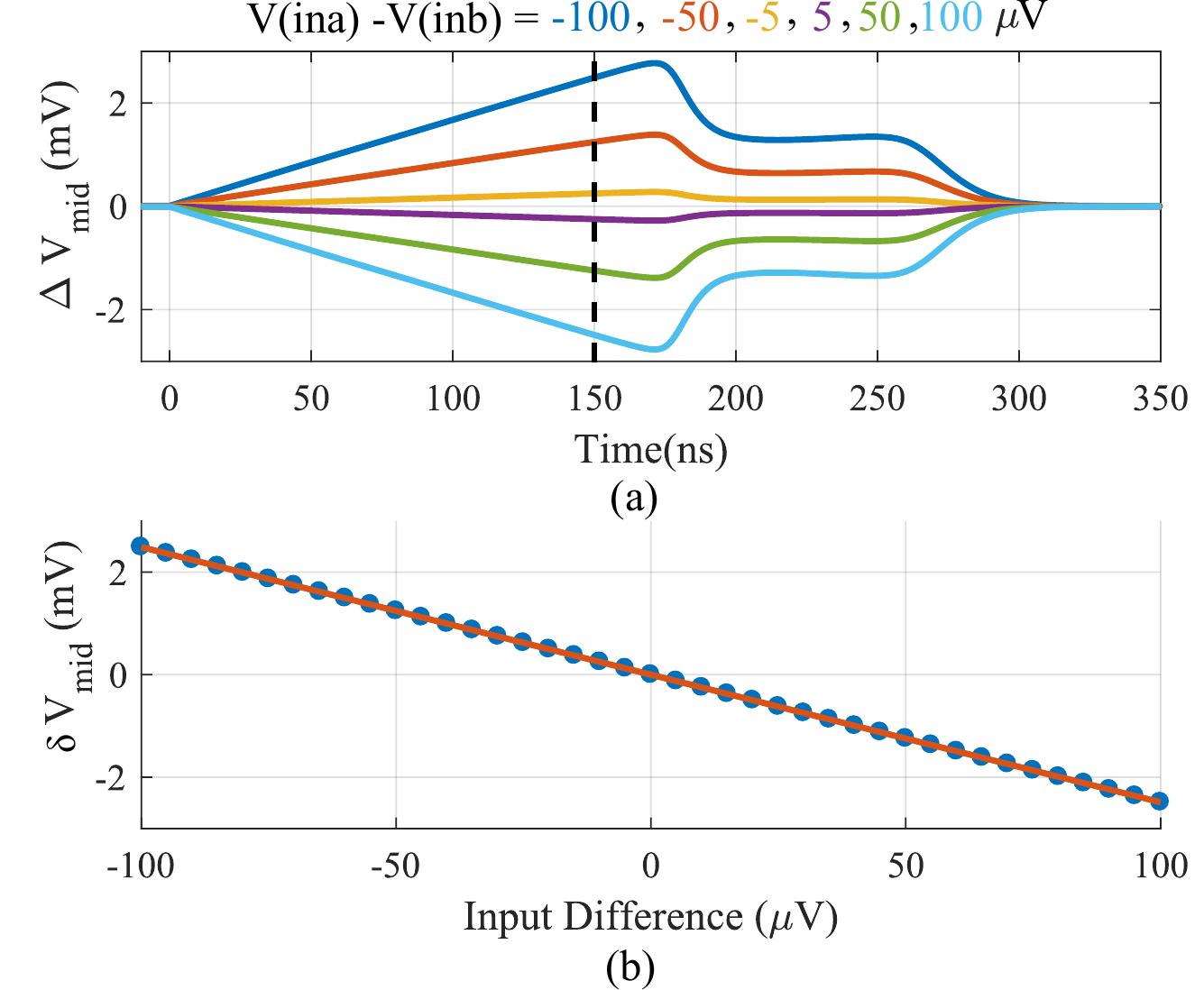}
\caption{Simulation results of differential gain analysis. In this simulation, $inb$ was set at $1.1V$ while $ina$ was swept from $1.1V - 100\mu V$ to $1.1V + 100\mu V$ with in steps of $5\mu V$. (a) $\Delta V_{mid}$ throughout an intentionally excessively long integration phase. As $V_{mida,b}$ increases the cascode transistors eventually triode causing the gain to peak and then start decreasing. Peak gain times occur at $t=170ns$ and are aligned within $1ps$ difference. An indicative integration time leaving substantial margin for error can be set to e.g. $150ns$ (dashed line in (a)). (b) Output voltage difference $\delta V_{mid}$ at integration time $\tau = 150ns$ vs. input differential voltage. The result is excellently fitted by a linear curve. The gain is constant at approx. $G=25$.}
\label{fig_4}
\end{figure}

The first experiment is illustrated in Fig. \ref{fig_4}(a). We observe that for all test inputs $\delta V_{in} \in \{-100, -50, -5, 5, 50, 100\}\mu V$ $\Delta V_{mid}$ increases linearly to a global peak at $\approx 170ns$ into the integration phase and then gradually decreases to zero, at which point both $V_{mida/b}$ have saturated and any potential difference they had is erased. The peak occurs because as we keep integrating, the voltage at $mida/b$ nodes eventually increases to the point where the cascode transistors enter the triode mode. This causes the rate of voltage accumulation on whichever $V_{mid}$ node is highest to slow first, allowing the other node to catch up (and leading to the post-peak drop in $\Delta V_{mid}$). At this point we are past maximum gain and continuing the integration eventually equalises the $V_{mid}$s.

Next, we note that the peak gain time is nearly perfectly aligned for all input samples; the maximum peak time difference is only $1ps$. The high quality of alignment arises because the time at which the $V_{mid}$ voltages start trioding the cascode transistors is determined primarily by the tail current and not the differential currents. The small discrepancy is explained by the fact that the peak gain time is technically determined by the time at which \emph{the first of $V_{mida/b}$} reaches the point where it triodes its cascode transistor. This has two key engineering implications: 1) It allows us to set a universally optimal DLC triggering time. 2) It states that the optimal trigger time is bounded by the trioding time obtained for $V_{mida}=\text{min}$ and $V_{midb}=\text{max}$ (or vice versa), in which case we have the fastest trioding corner.

The results from the 2nd experiment are shown in Fig. \ref{fig_4}(b). The differential output voltages $\delta V_{mid}$ for $\tau=150\,ns$ are plotted versus input differential voltage $\delta V_{in}$. We notice excellent gain linearity arising again from the extremely small effect that the differential voltages have on the behaviour of the voltages at $V_{mida,b}$. For this experiment the differential input voltage was swept on the basis of a fixed input $V_{midb} = 1.1V$ and a swept input $V_{inb} \in [1.1V - 100\mu V, 1.1V + 100\mu V]$. Results were linearly fitted yielding a gain of $G=25V/V$ ($28dB$) with excellent linearity throughout the range (MSE $= 0.0011$).

\subsubsection{Bandwidth}
\begin{figure}[!t]
\centering
\includegraphics [width=6cm]{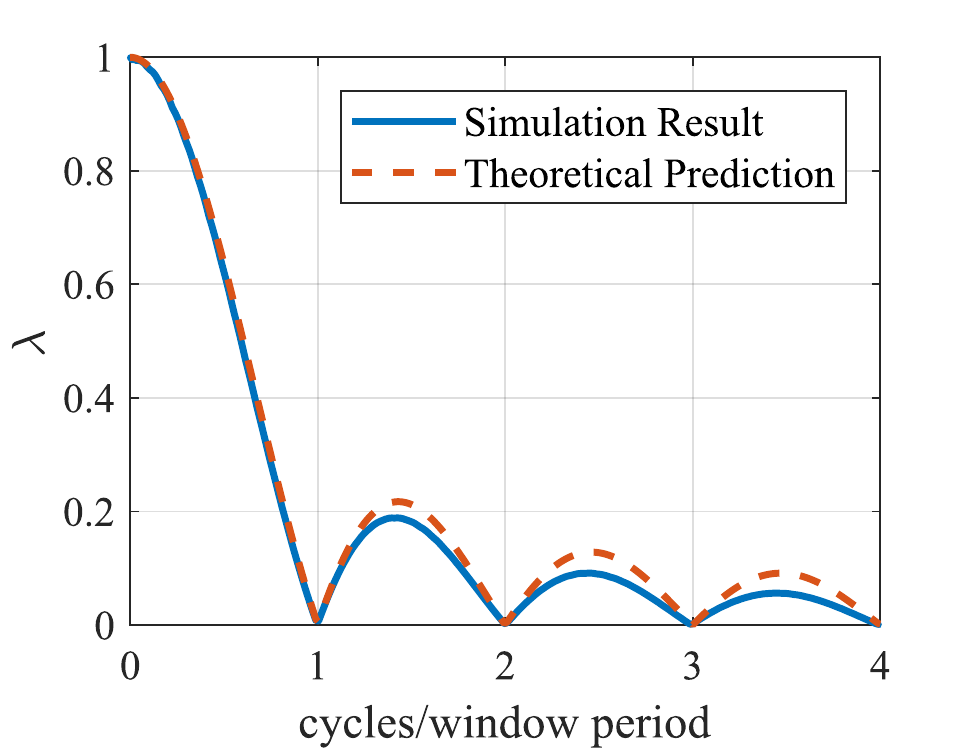}
\caption{Simulated bandwidth profile of proposed integrating amplifier. Attenuation factor as a function of tone frequency $\overline{f}$ in units of cycles/window period ($\frac{2\pi}{\tau} = \frac{\pi}{a}$), with $a = 150ns$ in this design. Dashed line indicates theoretical prediction.}
\label{BW_pic}
\end{figure}

We operated our amplifier with an integration period of $150ns$ as shown in Fig.\ref{fig_3} and ran a collection of transient analyses for fixed amplitude pure tone signal inputs. The tone frequencies ranged from 1Hz to $27MHz$ (covers around four cycles of window) and for each frequency the phases where stepped in increments of $10^o$. Additionally we also carried out a DC run ($\delta V_{in} = 100\mu V$). For each simulation run we looked at the amplifier output $\delta V_{mid}$ after $150ns$ of integration. The outcome was a plot of maximum $|\delta V_{mid}|$ as a function of frequency, as illustrated in Fig.\ref{BW_pic} (normalised to $|\delta V_{mid}|$ at DC). The resulting curve is closely bounded by the envelope calculated by Eq. \ref{lambdaf} indicating no surprises. To keep $\lambda > 20\%$, the bandwidth achieves four fifths cycles/window period in Fig.\ref{BW_pic} that yields $5.4MHz$ bandwidth.

We note that if we assume that the highest frequency spectral component of interest in a neural spike lies at $10 k$Hz, the maximum attenuation of this particular design is around $0.08\%$. Therefore we can reliably sample spiking waveforms with this design.


\subsubsection {Tuneable Range and Tuning Sensitivity}
To obtain the tuneable range and sensitivity of implanted memristive devices, multiple transient simulations such as those seen in Fig.\ref{fig_2} can be repeated while sweeping both RRAM device resistive states ($R1$ and $R2$). By tracking at what difference $\Delta V_{in}$ the outputs flip value we can obtain an estimate for the offset. The quality of the estimate is calculated as follows: if at cycle $n$ we had $V_{outa}=0$ and at cycle $n+1$ we obtained $V_{outa}=1$, it means that somewhere between $\delta V_{in}|n$ and $\delta V_{in}|(n+1)$ we crossed the amplifier's offset voltage. The tracking will be applied in both ascending and descending phase, after which offset voltage will be averaged. Assuming that the amplifier always makes a decision at approximately the same relative time in each cycle (in our case always at $150ns$ into the integration phase), the duration of this interval is fixed and given by the total swept range over the number of sampling cycles. In our case, we run 200 cycles ($10 \mu s$/cycle for a total duration of $2ms$) and sweep the input across a range of $400 \mu V$ ($200 \mu V$ ascending and $200 \mu V$ descending)

Table \ref{table_Vos} shows the offset voltage as a function of $R1, R2$. From there we observe: 1) The overall trimming range for this particular design is $\approx 235\mu V$. 2) The maximum induced offset occurs, as expected, at the maximum $R1, R2$ imbalance corners. 3) The offset sensitivity is close to $1 \mu V/k\Omega$ for any combination of $R1, R2$. 4) The table is almost symmetric (as expected). The slight asymmetry indicates that the common mode voltage influences the offset voltage. This effect will be the subject of a dedicated study. Finally, the quoted offsets were checked and are the same both on the upward and the downward slopes, indicating no history-dependence.



\begin{table}[!h]
\centering
\caption{Offset voltage of pre-amplifier vs. RRAM device resistive state quoted at $5\mu V$ resolution.}
\begin{center}
 \begin{tabular}{|c|c|c|c|c|c|} 
 \hline
 R1$\backslash$R2 & 10$k\Omega$ & 40$k\Omega$ & 70$k\Omega$ & 100$k\Omega$ & 130$k\Omega$ \\ 
 \hline
 10$k\Omega$ & 0 & 35 & 60 & 90 & 120 \\ 
 \hline
 40$k\Omega$ & -35 & 0 & 25 & 55 & 95 \\
 \hline
 70$k\Omega$ & -60 & -25 & 0 & 25 & 55 \\
 \hline
 100$k\Omega$ & -85 & -50 & -25 & 0 & 30 \\
 \hline
 130$k\Omega$ & -115 & -80 & -50 & -25 & 0 \\
 \hline
\end{tabular}
\end{center}
\label{table_Vos}
\end{table}

\subsubsection{Input-Referred Noise}
\begin{figure}[!t]
\centering
\includegraphics [width=8cm]{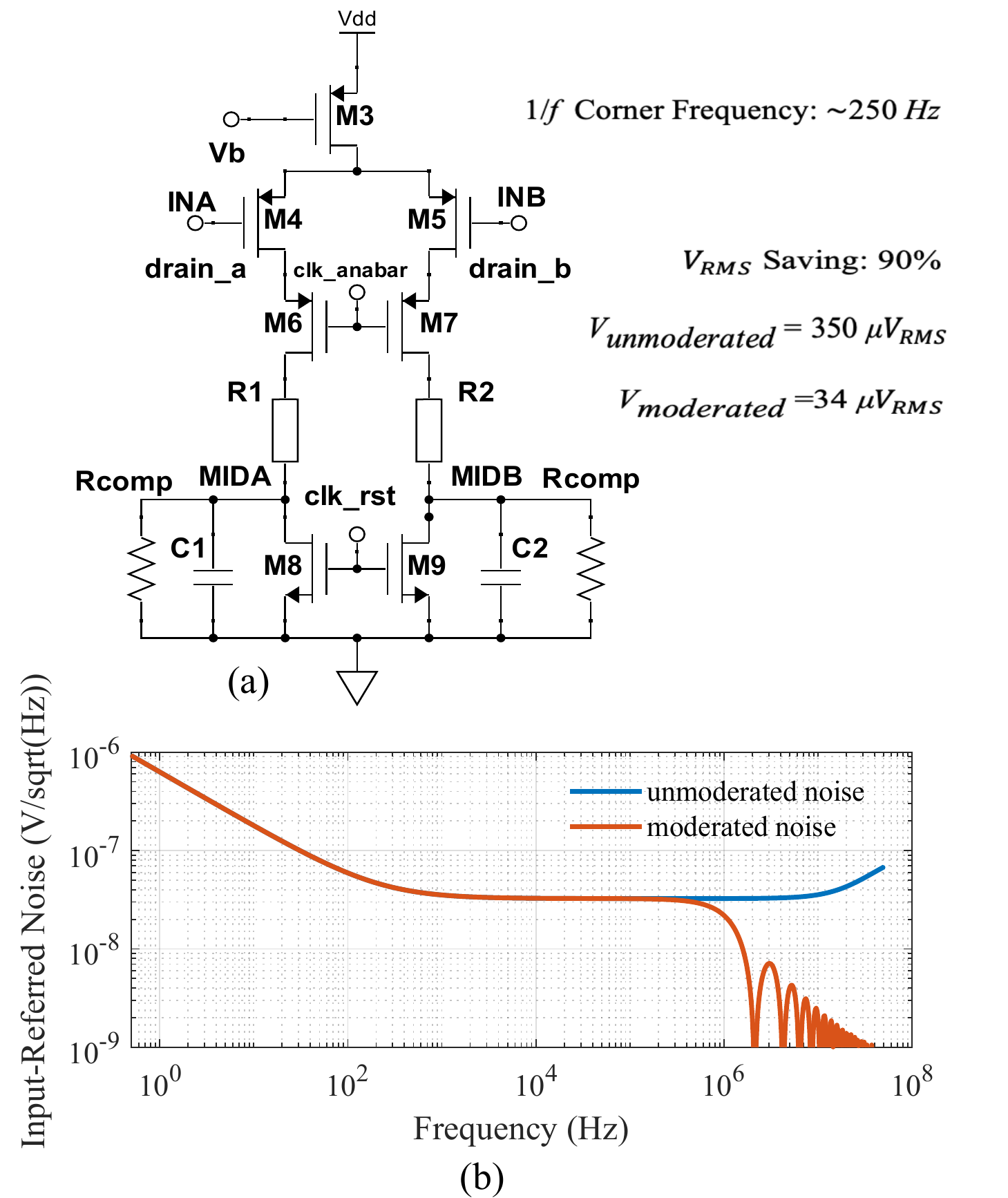}
\caption{Noise Simulation. (a) Schematic used for running noise analysis. Noiseless baseline current compensation resistors $R_{comp} = 330k\Omega$ were used. The resistors divert the baseline current coming from the tail transistor M3 so that at equilibrium any remaining voltage fluctuations on nodes MIDA,B are attributable to noise. (b) The noise spectrum presents unmoderated and moderated input-reffer noise respectively. }
\label{noisesch}
\end{figure}

To estimate the noise behaviour we employ the following trick: we take the core of the basic circuit shown in Fig. \ref{fig_1}, balance the inputs and add a pair of ideal, noiseless resistors that sink the baseline value of $i_{tail/2}$ for some suitably chosen equilibrium voltage $V_{mida/b} = V_{equil}$ within the amplifier's integrating range. This is shown in Fig. \ref{noisesch}(a) (note that we have removed M17\&M18 for simplicity - they only increment node capacitance by a small fraction). Then, we need to run our noise analysis and apply the sinc moderation (Eq. \ref{lambdaf}) in order to obtain our final results.

Before we begin, we need to make some key observations/assumptions: 1) At DC equilibrium, what is left on $V_{mid}$ after removing the baseline tail currents is fluctuations due to noise; there is no other possible source of fluctuation. 2) Any distortions introduced by the finite impedance of the compensation resistors is negligible due to the minute input signals at play. 3) Input-referred output noise levels are expected to be comparable throughout the entire integration range given that most of the noise is generated by the input differential pair. Additionally, running the noise test at half-gain is compensation against underestimating the noise generated by other sources (most notably the cascode pair). Now we can run our noise analysis.

For baseline compensation resistances $R_{comp} = 330k\Omega$, we get $V_{equil} \approx 0.5V$ and a noise spectrum (with and without sinc moderation) as shown in Fig. \ref{noisesch}(b). Across a $[0.05Hz-50MHz]$ bandwidth we obtain a root-mean square (RMS) voltage noise level of $\approx 350 \mu V$ unmoderated, dropping to $34\mu V$ moderated. This represents a saving of $\approx 90\%$. We also observe a $1/f$ corner frequency around $250Hz$. We have tested that expanding the included noise bandwidth both to the left and to the right does not change the above figures significantly. The present analysis excludes noise contributions from the RRAM devices.

The overall result suggests that for neural probing, the noise levels obtained for this design may still be slightly too high, especially if we include additional noise from the RRAM devices. In this case switching to longer integration periods would help.


\subsection{Common Mode Effects}
\subsubsection{Input and Range}
\begin{figure}[!t]
\centering
\includegraphics [width=8cm]{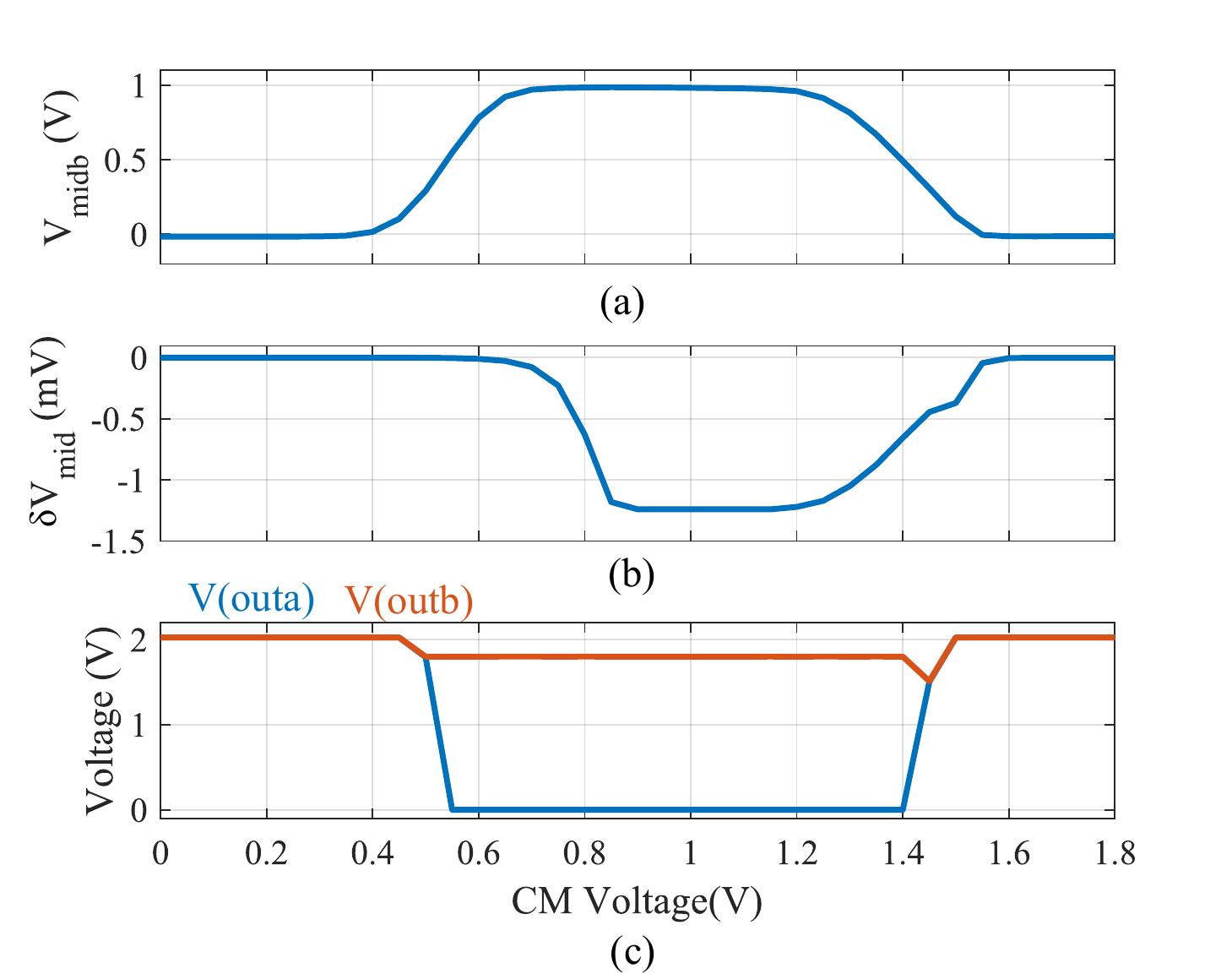}
\caption{Input range results of pre-amplifier. In this simulation, the common mode voltage was swept from zero to $1.8V$ with $50\mu V$ differential input. For $V_{CM} \in [0.5-1.4]V$ we notice that $V_{midb}$ (a) reaches sufficiently high voltage to prompt a stable output from the DLC (c) within $50ns$ of triggering, and for our chosen differential input the output is always correct. However, the analogue gain in (b) of the core is maximised in the narrower range $[0.9, 1.3]V$.
}
\label{fig_5}
\end{figure}

In order to experimentally demonstrate the input range of the amplifier we performed a series of experiments querying different potential range limitation factors in practice. First, we checked the behaviour of the system at different stages as a function of common mode voltage by running a series of integration cycles whilst sweeping $V_{CM}$ from 0V to VDD in steps of $50 mV$. At each run the differential input was $50 \mu V$ and the outputs were registered after integrating for $150 ns$. Results were registered at: i) $V_{midb}$, ii) $\delta V_{mid}$ and iii) the overall system output after the DLC. Results are shown in Fig. \ref{fig_5}. Note: in order to check for possible input signal history-dependence during these tests, each test integration cycle was preceded by three integration cycles ran with $V_{CM}=1.8V$. We have sample-tested a few runs with initial $V_{CM}$ between 0.1V and 1.8V and confirm that the history-dependence effect is negligible.

From the results in Fig. \ref{fig_5} we can draw three key conclusions: 1) The DLC successfully triggers for $V_{CM}$ between approx. $0.5V$ and $1.4V$. This means that $V_{midb}$ is sufficiently high for the DLC to settle to an output within $50 ns$ of it triggering (which occurs when $clk$ goes high). 2) In this case the DLC provides the correct answer so long as it triggers, but this might change towards the edges of the range once we take noise into account. 3)  The actual analogue gain of the amplifier remains close to maximum ($\approx 28dB$) within a narrower region: $approx [0.9, 1.3]V$. We would recommend that maximum gain area is taken as the effective $V_{CM}$ range in order to maximise the chances of correctly capturing small differential inputs under noisy conditions. Nevertheless, this shows that by de-rating the specification of the amplifier to higher $\delta V_{mid}$ we can extend its effective input range.

\begin{figure}[!t]
\centering
\includegraphics [width=8cm]{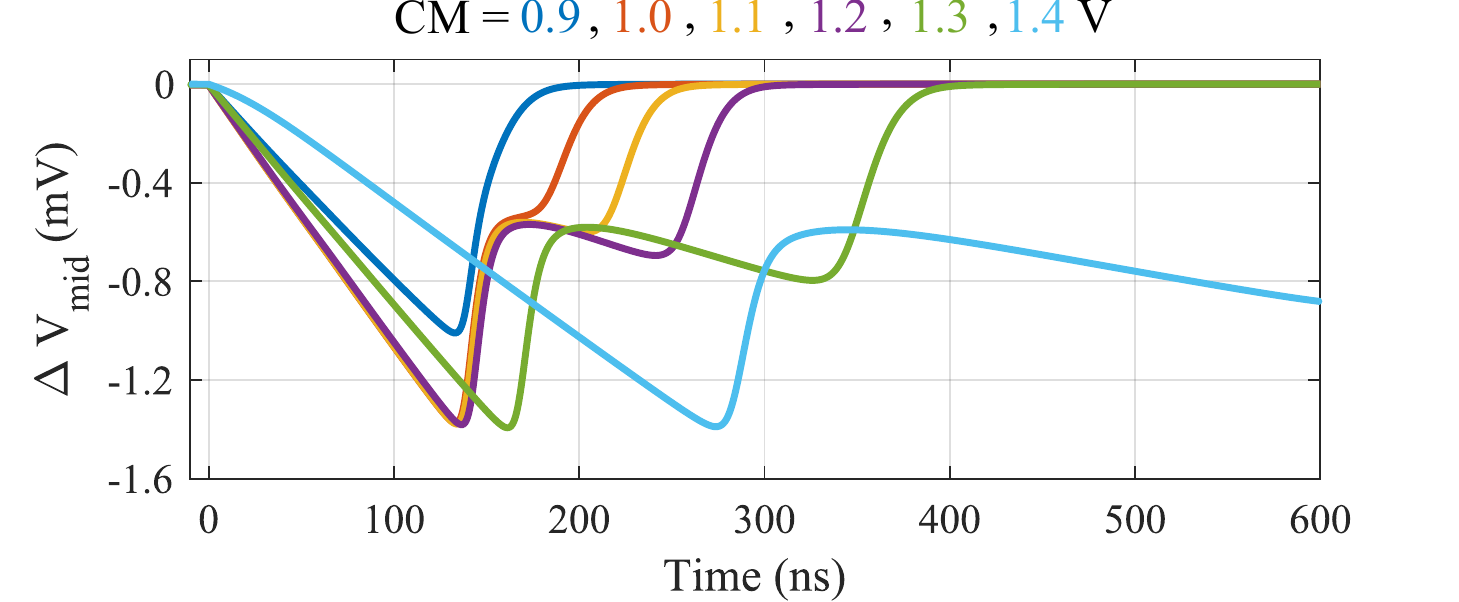}
\caption{Intermediate differential output $\Delta V_{mid}$ evolution as a function of $V_{CM}$. Differential input voltage is $50\mu V$ and the integration phase is not time-constrained, (see Fig.\ref{fig_4}). Voltage traces for different $V_{CM}$s follow each other closely except in the edge cases $V_{CM} \in \{0.9V, 1.4V\}$.}
\label{fig_6}
\end{figure}

In order to visualise the effects leading to loss of gain outside the region $V_{CM}\in[0.9, 1.3]V$ we ran some unrestricted integration tests as shown in fig. \ref{fig_4} for different values of $V_{CM}$. The results are shown in Fig. \ref{fig_6} where we observe that for $V_{CM}$ between $1.0V$ and $1.3V$ the integration traces follow each other very closely, with traces at $0.9V$ and $1.4V$ beginning to show more substantial deviations. We note how excessively low $V_{CM}$s shorten the peak without shifting (a result of desaturating the input differential pair but not changing the integration range) whilst excessively high $V_{CM}$s shift the peak without changing its magnitude.

\subsubsection{CMRR and CMGD}
For evaluating the CMRR we set the differential input to 0V and swept $V_{CM}$ between $[0.9, 1.4]$V. Since we deliberately don't account for process variations and mismatch in this work, we obtain the expected common mode gain of 0.

\begin{figure}[!t]
\centering
\includegraphics [width=8cm]{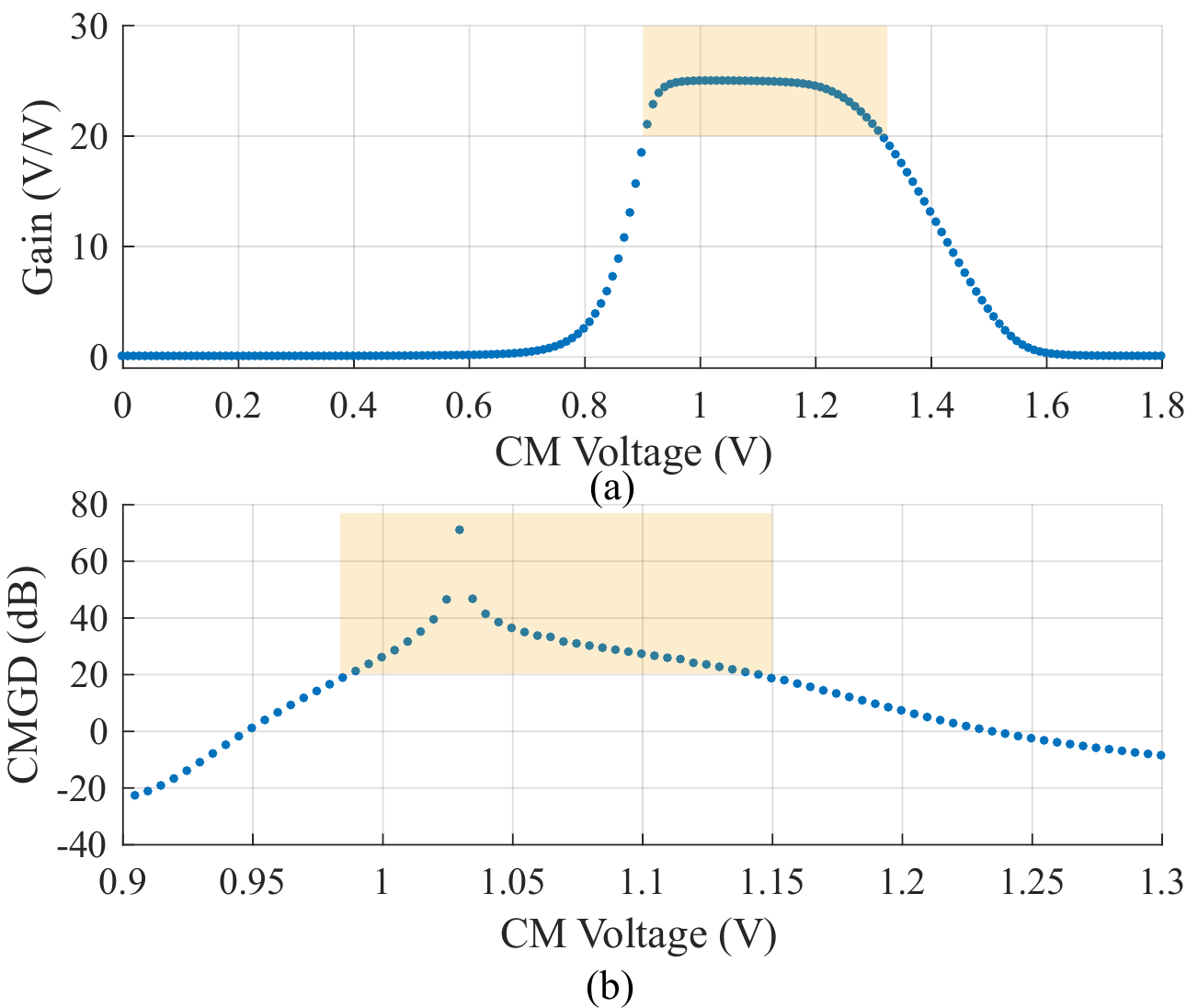}
\caption{CMGD simulation results. (a) Core amplifier gain vs. $V_{CM}$ ($150ns$ integration period). The gain remains high and stable in the highlighted area ($V \in [0.99V, 1.14V]$). (b) CMGD appropriately converted to dB for the range highlight area in (a). New highlight indicates CMGD$\geq 20dB$.}
\label{fig_7}
\end{figure}

For CMGD, we run a series of integration runs with fixed differential input voltage ($50 \mu V$) and sweep $V_{CM}$ in steps $10mV$ and plot the gain as illustrated in Fig. \ref{fig_7}(a). The highlighted region where the gain maximises is then resampled at $5mV$ step and for each consecutive pair of data points we calculate the derivative. As per Eq.\ref{unfold1} this yields our CMGD. Converting appropriately we obtain CMGD$\geq 20dB$ for $V_{CM} \in [0.99V, 1.14V]$. To exemplify this effect, a 0.15V change in common mode voltage $\Delta V_{CM}$ causes less than 1.5\% change in the output of the amplifier core ($\frac{dG}{dV_{CM}}\Delta V_{CM}$).

\subsection{Power Consumption}
The power consumption has to be assessed for all operating phases of the pre-amplifier. The most power-hungry phase is the reset phase since it is the only one where a DC path exists between the power supplies. For this reason the reset phase should be kept as short as possible. However, it is also during the reset phase that the core amplifier reaches steady state at all nodes so that the integrating phase can then commence without any history-dependence, i.e. influence from or `memory of' its previous inputs. Finding the optimal reset phase duration is a key optimisation task for this design. Next, the cost associated with the integration and digitisation phases can be split into two main components: First, the integration cost is equal to charging the core amplifier's capacitors from GND to their equilibrium level, where the integration self-terminates ($\approx 1.26V$ in our case - note how this integration cost currently spans both integration and digitisation phases because we do not stop the integration once we trigger the DLCs). Second, the comparison cost is equal to the energy needed to operate the DLC. Finally, during the `off' phase power dissipation is mainly down to leakages.

Through one detection cycle ($350ns$), the average energy consumption is $1.927 pJ$, of which $663fJ$ during the reset phase, $814fJ$ during the integration phase and $450fJ$ during digitalisation. This yields a power rating of $5.5\mu W$ for continuous operation (no off phase), of which the core amplifier accounts for $5\mu W$. If we operate the amplifier at typical biointerface sampling rates of $\approx 20k$Hz, power dissipation becomes $38.5nW$ (assuming practically zero `off' mode dissipation).

For a more complete, multiple channel pre-amplifier, additional power will be dissipated by 1) the current reference generation unit (III in Fig.\ref{fig_1}), 2) the control system, including $clk\_ana$, $clk\_anabar$, $clk\_rst$ and $clk$ generators. Both of the above would be shared across multiple channels, yielding a certain degree of amortisation.



\section{Discussion}\label{secDis}
From the analysis and simulation  of the integrating amplifier we highlight some key conclusions:

First, the performance improvement of the integrating amplifier over more traditional e.g. Harrison designs relies on the integration process, which enhances the gain and decreases the effective bandwidth (helping reduce noise in the process). To visualise this let us consider an integrating amplifier using the same tail current as a standard OTA first stage. During integration the power dissipation is effectively the same, but the gain and bandwidth are different. In this sense the design represents a trade-off between gain and bandwidth without changing power dissipation or using feedback.

Next, we note that there is a natural trade-off between tail current and integration time while keeping the overall energy dissipation approximately constant. This is the result of the fixed duration of the reset phase (just enough to clear any residual charge at the $V_{mid}$ nodes) and the fact that energy consumption during the integration phase only depends on the size of the load caps and the voltage change across them during that phase. Thus, in principle we can design for a wide range of required sampling rates or bandwidths for the same energy budget.

The trade is not completely free: Changing the tail current affects gain, bandwidth and noise performance, by altering the $g_m$s of all transistors involved and the integration period. Furthermore, if using real RRAM devices with non-linear IV curves, changing the tail current also changes the static resistance of the RRAM devices. Together with changes in transistor $g_m$s this means that the tuneability range is also affected since it depends on the impedance balance between RRAM and transistors. Thus, whilst the integrating amplifier clearly offers a lot of design flexibility, the precise design trade-off space is also not trivial, much like as it is for OpAmps. This is an important subject meriting its own dedicated study.

The last design decision to highlight concerns the size of the load capacitors $C$. The gain analysis in section \ref{secBack} shows that $C$ doesn't affect the gain, but it does affect the integration period and therefore can be used to adjust the bandwidth, if for some reason that cannot be achieved by tweaking the tail current. Effectively it is a design parameter that trades away energy for design flexibility.
 
In terms of operation, we note the importance of ensuring that the integrating amplifier is cleared properly in preparation for each integrating phase in order to avoid history-dependence of the output. This means that all node voltages should be equalised across the left and right branches prior to the commencement of the sensitive integration phase. In the current design this is achieved by forcefully flushing the system during the reset phase, but more energy-efficient approaches are under development as the rest phase represents a substantial fraction of the energy budget.

Finally, we compare our amplifier's performance with a few standard designs as shown in Table \ref{table_Comp}. We observe a slightly reduced gain and increased noise levels traded against power dissipation as a result of our design decisions so far. Importantly, for relatively low precision operations such as threshold detection of neuronal spikes a 10-fold increase in noise may be an acceptable price for a 100-fold reduction in power dissipation. We also note that the present design is not completely optimised, with an increase in integration time as a very promising avenue of investigation for decreasing noise levels within the same power envelope.

\begin{table}[h!]
\centering
\caption{Performance and comparison of the proposed amplifier. IRN: input-referred noise}
\begin{center}
 \begin{tabular}{c|c|c|c|c} 
 \hline
 Work & \cite{Harrison2003} & \cite{8745479} & \cite{9153020} & This work\\
 \hline
 \hline
 Tech. ($\mu m$) & 1.5 & 0.18 & 0.18 & 0.18\\
 \hline
 Power ($W$) & $40\mu$ & $3.24\mu$ & $1.5\mu$ & $38.5n @ 20kHz$\\
 \hline
 Gain ($dB$) & 40 & 40 & 60 & 28 \\
 \hline
 BW ($Hz$) & $7.5k$ & $5.4k$ & $10k$ & $5.4M$ \\
 \hline
 \multirow{2}{5em}{IRN @freq. ($\mu V_{rms}$)} & 2.1  & 2.14  & 3.4  & 34  \\
 & $0.5-50k$ & $200-5k$ & $0.5-10k$ & $0.5-50M$\\
 \hline
\end{tabular}
 \label{table_Comp}
\end{center}
\end{table}

\vspace*{-1\baselineskip}

\section{Conclusion}\label{secConc}
In this work we have performed a theoretical analysis of the core functionality of memristive integrating amplifiers and used industrial CAD-level simulations to provide a specific example for an integrating amplifier design targeting electrophysiological applications. Throughout our analysis we have concluded that the performance enhancement over traditional, continuous mode amplifiers can be most intuitively understood as a gain boosting effect arising from the integration process and showed how this process erodes the amplifier's effective bandwidth (which is desirable for electrophysiology applications). Moreover, we have explained how standard metrics of amplifier performance such as gain and input common mode range, but also new metrics such as offset voltage tuneability range can be described by governing equations for use by designers. Finally, we implemented an exemplar design in commercially available $180nm$ CMOS and demonstrated typical values for all studied performance parameters that can be expected from a $0.18 \mu m$ node technology. These included gain of $25V/V$, offset tuning range of $235 \mu V$, input-referred noise of $34 \mu V_{rms}$ and power dissipation of $38.5 nW$ at $20 k$Hz sampling rate. These are competitive vs current literature for an not fully optimised design.

This work is a stepping stone towards de-risking and documenting the RRAM-based integrating amplifier. We believe that the trade-off induced by the integration process in combination with the offset trimming enabled by RRAM has the potential to add a powerful circuit topology to the arsenal of the analogue designer.

\section*{Acknowledgment}

The authors would like to acknowledge this work was supported in part by the Royal Society Industry Fellow PhD Student Scholarship and Engineering and Physical Sciences Research Council (EPSRC) under Grant EP/R024642/1 in Functional Oxide Reconfigurable Technologies (FORTE) programme.

\ifCLASSOPTIONcaptionsoff
  \newpage
\fi



\bibliographystyle{IEEEtran}
\bibliography{ref}

%

\vspace*{-4\baselineskip}

\begin{IEEEbiography}
[{\includegraphics[width=1in,height=1.25in,clip,keepaspectratio]{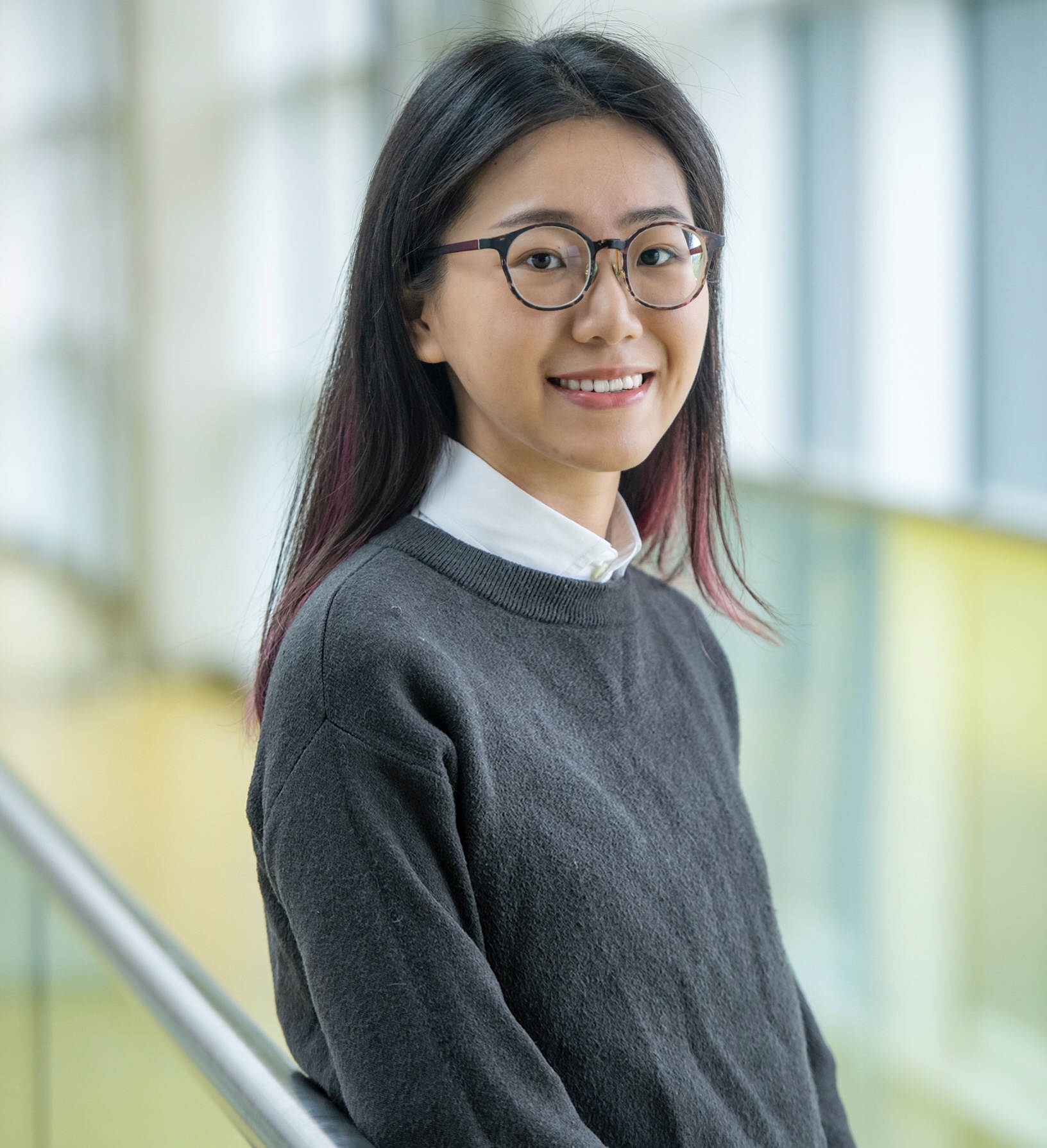}}]{Jiaqi Wang}
received her bachelor degree in Microelectronic Science and Engineering from Shenzhen University, China, in 2017 and her M.Sc. degree in Microelectronics Systems Design from University of Southampton, UK, in 2018. And she is currently pursuing her PhD studies in Zepler Institute, University of Southampton, working towards memristor-based hardware design, analogue and mixed-signal integrated circuit design for biosignal processing.
\end{IEEEbiography}

\vspace*{-3\baselineskip}

\begin{IEEEbiography}
[{\includegraphics[width=1in,height=1.25in,clip,keepaspectratio]{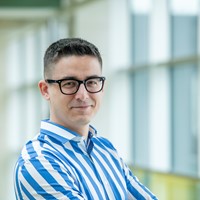}}]{Alexander Serb}
received his degree in Biomedical Engineering from Imperial College in 2009 and his PhD in Electrical and Electronics Engineering from Imperial College in 2013. Currently he is a research fellow at the Zepler Institute (ZI) dept., University of Southampton, UK. His research interests are: cognitive computing, neuro-inspired engineering, algorithms and applications using RRAM, RRAM device modelling and instrumentation design.
\end{IEEEbiography}


\vspace*{-2\baselineskip}

\begin{IEEEbiography}
[{\includegraphics[width=1in,height=1.25in,clip,keepaspectratio]{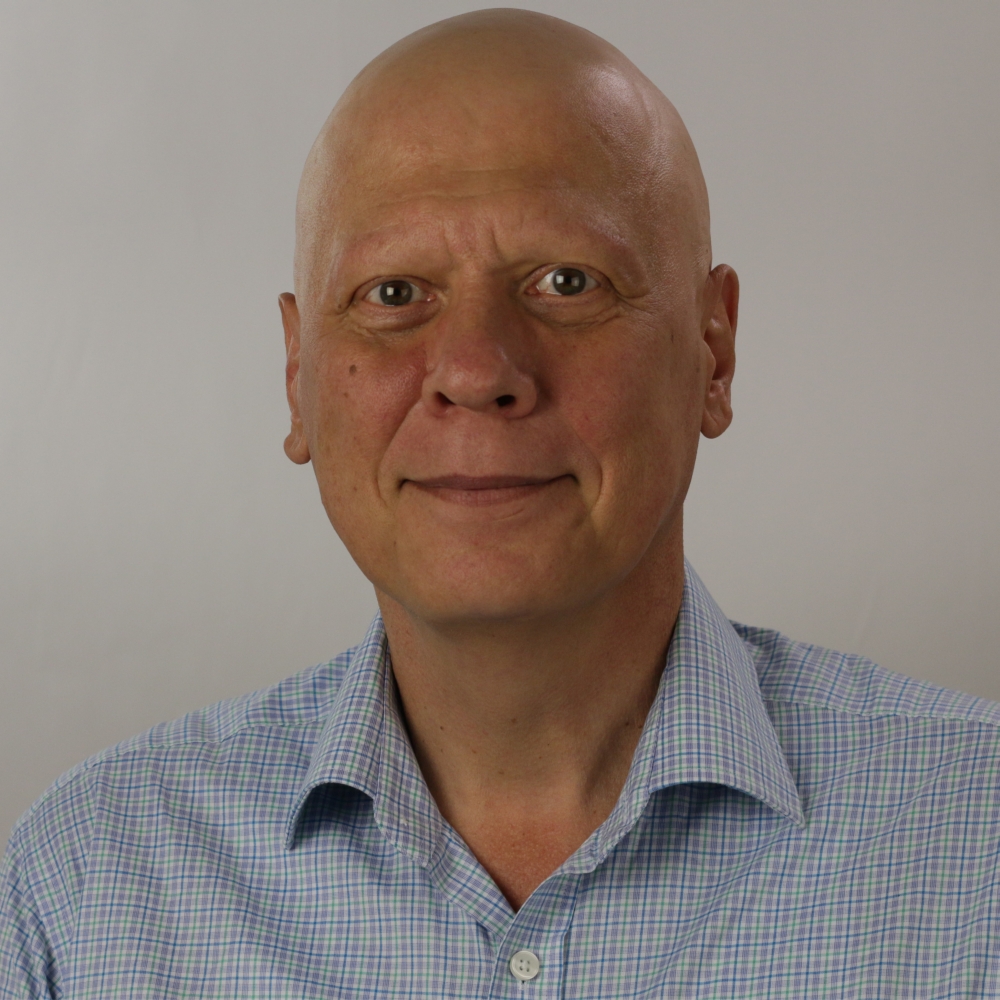}}]{Christos Papavassiliou}
received the B.Sc. degree in physics from the Massachusetts Institute of Technology, and the Ph.D. degree in applied physics from Yale University.,He is currently with the Electrical Engineering Department, Imperial College London. He currently works on memristor applications, sensor devices, and systems and antenna array technology. He has contributed to over 70 publications on weak localization, GaAs MMICs, and RFIC.
\end{IEEEbiography}

\vspace*{-2\baselineskip}

\begin{IEEEbiography}
[{\includegraphics[width=1in,height=1.25in,clip,keepaspectratio]{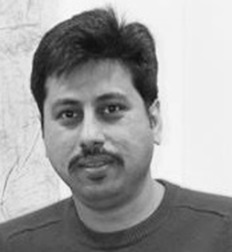}}]{Sachin Maheshwari}
received his Bachelor's degree in Electrical and Electronic Engineering from the ICFAI University, India and Master's in Microelectronics from Birla Institute of Technology and Science, Pilani, India. He then obtained his PhD degree in Electronics Engineering from the University of Westminster, London, U.K. Currently, he is a Research Fellow at the Centre of Electronics Frontiers, University of Southampton, Southampton, U.K. His research interest is in Energy Recovery Logic and Regenerative Neural Networks.
\end{IEEEbiography}

\begin{IEEEbiography}
[{\includegraphics[width=1in,height=1.25in,clip,keepaspectratio]{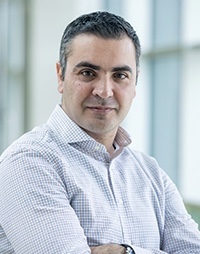}}]{Themistoklis Prodromakis}
received his Bachelor in Electrical and Electronic Engineering from the Department in electrical and Electronic Engineering, University of Lincoln, UK. He then obtained his MSc degree in Microelectronics
and Telecommunications from the Department of
Electrical Engineering and Electronics, University
of Liverpool, UK while his PhD in Electrical
and Electronic Engineering was obtained from
the Department of Electrical and Electronic
Engineering, Imperial College London. He held a
Corrigan Fellowship in Nanoscale Technology and Science with the Centre for Bio-inspired Technology, Imperial College London, London, U.K., and a Lindemann Trust Visiting Fellowship with EECS UC Berkeley, Berkeley,
CA, USA. He is a Professor of nanotechnology and EPSRC and Royal Society Industry Fellow affiliated with the Southampton Nanofabrication Centre, University of Southampton, Southampton, U.K. His background is
in electron devices and nanofabrication techniques. His current research interests include bio-inspired devices for advanced computing architectures and biomedical applications. Prof. Prodromakis is a Fellow of the IET and the Institute of Physics.
\end{IEEEbiography}

\balance

\vfill


\end{document}